%% file: main.tex
\documentclass[letterpaper,twocolumn,10pt]{article}
\usepackage{usenix2019_v3}

\usepackage{amsmath}

\usepackage{filecontents}

\usepackage{amssymb}
\usepackage{inconsolata}
\usepackage{array}
\usepackage{booktabs}
\usepackage{fullpage}
\usepackage{threeparttable}
\usepackage{wasysym}
\usepackage{flushend}
\usepackage{caption}
\usepackage{subcaption}
\usepackage{longtable}
\usepackage{scrtime}
\usepackage[multiple]{footmisc}
\usepackage{footnote}
\usepackage{makecell}
\usepackage{multirow}
\usepackage{rotating}
\usepackage{multicol}
\usepackage{graphicx}
\usepackage{float}
\usepackage{listings}
\usepackage{cleveref}
\usepackage{textcomp}
\usepackage{enumitem}
\usepackage{tikz}
\usepackage{verbatim}
\usepackage{booktabs}
\usepackage{siunitx}
\usepackage{pifont}
\usepackage{microtype}
\usepackage[compact]{titlesec}
\usepackage[toc,page]{appendix}

\crefformat{section}{\S#2#1#3} 
\crefformat{subsection}{\S#2#1#3}
\crefformat{subsubsection}{\S#2#1#3}

\usepackage{epsfig,hyperref,color,graphicx,xspace,pifont,dirtree}
\hypersetup{%
  linkcolor=black
}

\usepackage{fancyhdr}
\usetikzlibrary{shapes.geometric, arrows}
\usepackage{pgfplotstable}

\usepackage{pgfplots}
    \pgfplotsset{compat=1.3}
    
    \definecolor{bblue}{HTML}{4F81BD}
    \definecolor{rred}{HTML}{C0504D}
    \definecolor{ggreen}{HTML}{9BBB59}

\newcommand{\tpara}[1]{\paragraph{#1}}
\newcommand{\rom}[1]{{\em\lowercase\expandafter{(\romannumeral #1\relax)}}}
\newcommand{\nom}[1]{{\em\lowercase\expandafter{(#1\relax)}}}

\newcommand\cve[1]{{\href{https://cve.mitre.org/cgi-bin/cvename.cgi?name=CVE-#1}{CVE-#1}}}

\definecolor {processblue}{cmyk}{0.80,0,0.20,0.30}
\definecolor{darkgreen}{rgb}{0.18,0.54,0.34}
\definecolor{maroon}{rgb}{0.64,0.16,0.16}
\definecolor{darkpink}{rgb}{0.75,0.25,0.5}

\lstdefinestyle{customc}{
   breaklines=true,
  frame=ltrb,
  language=C,
  basicstyle=\ttfamily\scriptsize,
  showstringspaces=false,
  keywordstyle=\color{blue}\bfseries,,
  identifierstyle=\color{processblue}\bfseries,,
  stringstyle=\color{processblue}\bfseries,,
  commentstyle=\itshape\color{darkgreen}\bfseries,,
    tabsize=2,
    morekeywords={include, clone, vdom_malloc, s_access_disable,s_clone, s_access_enable,vdom_free,imc_kill,imc_grant,imc_create,SLABEL,MEMDOM_READ, MEMDOM_WRITE,DECLARE_BITMAP}
}
\begin{document}

\date{}

\title{\large \bf Enclave-Aware Compartmentalization and Secure Sharing with Sirius}

\author{
{\rm Zahra Tarkhani}\\
University of Cambridge
\and
{\rm Anil Madhavapeddy}\\
University of Cambridge
} 

\maketitle

\begin{abstract}
Hardware-assisted trusted execution environments (TEEs) are critical building blocks of many modern applications. However, they have a one-way isolation model that introduces a semantic gap between a TEE and its outside world. This lack of information causes an ever-increasing set of attacks on TEE-enabled applications that exploit various insecure interactions with the host OSs, applications, or other enclaves.
We introduce Sirius, the first compartmentalization framework that achieves strong isolation and secure sharing in TEE-assisted applications by controlling the dataflows within primary kernel objects (e.g. threads, processes, address spaces, files, sockets, pipes) in both the secure and normal worlds.
 Sirius replaces ad-hoc interactions in current TEE systems with a principled approach that adds strong inter- and intra-address space isolation and effectively eliminates a wide range of attacks. We evaluate Sirius on ARM platforms and find that it is lightweight ($\approx 15K$ LoC) and only adds $\approx 10.8\%$ overhead to enable TEE support on applications such as httpd, and improves the performance of existing TEE-enabled applications such as
 the Darknet ML framework and ARM's LibDDSSec by $0.05\%-5.6\%$.
\end{abstract}

\input{intro.tex}

\input{back}

\input{overview.tex}

\input{design.tex}

\input{implementation.tex}
\input{eval.tex}

\input{related.tex}

\input{discussion.tex}
\input{conc.tex}

\input{appendix.tex}


\bibliographystyle{plain}
\bibliography{references}

\end{document}

%% file: intro.tex
\section{Introduction }\label{intro}

 


 
Hardware-assisted trusted computing primitives such as ARM TrustZone~\cite{arm2012architecture}, Intel SGX~\cite{intel2019linuxsgx}, AMD SEV~\cite{morbitzer2018severed} or RISC-V Keystone~\cite{dayeol2019keystone} exist to establish strong security guarantees even in the presence of malicious privileged code. These TEEs\footnote{For simplicity, we use the terms TEE and enclave interchangeably.} assume a threat model in which only the CPU itself is trusted, and not the host applications, OS, or hypervisor. They expect in-enclave code to be small, verifiable, and need minimal external interactions~\cite{ferraiuolo2017komodo}. 

\begin{figure}[t]
\centering
\includegraphics[width=\linewidth]{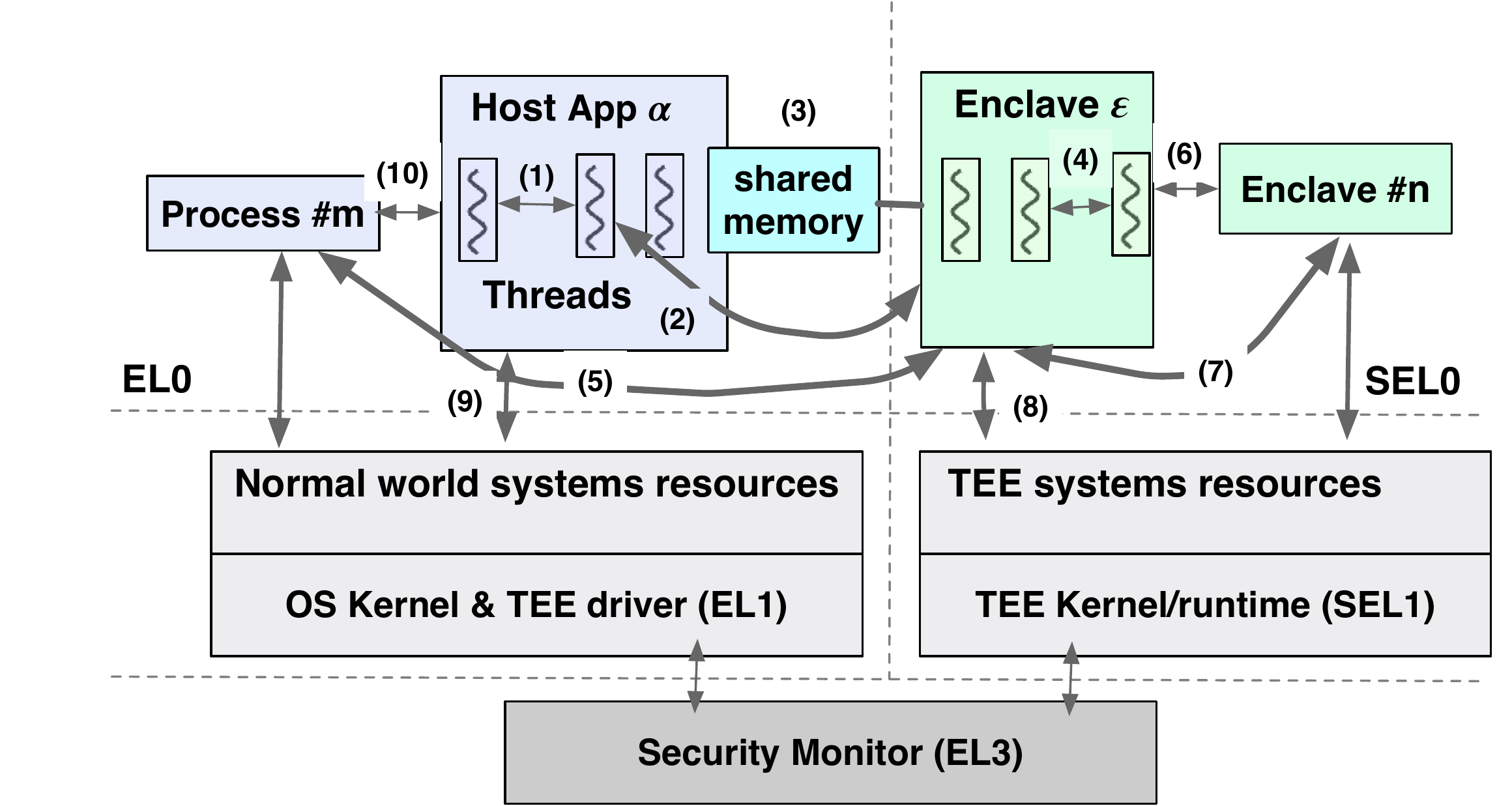}
\caption{Simplified data flows in TZ-based TEE systems}
\label{fig:risks}
\end{figure}

However, in practice, TEEs are used in much more complex application architectures like
web services~\cite{aublin2017talos,kim2018sgx,hunt2016ryoan}, secure payments~\cite{trusty,solutions2013white,keystore,apple}, databases~\cite{priebe2018enclavedb}, autonomous vehicle control~\cite{dds}, and privacy-preserving machine learning~\cite{hunt2016ryoan,grover2018privado,ohrimenko2016oblivious,schuster2015vc3,tramer2018slalom,hunt2018chiron}. 
These applications need fast bidirectional communications with their enclaves (e.g. via shared memory or RPC) and rely on OS facilities for multithreading, networking, file operations, and IPC with other processes or enclaves.
Despite the fact that TEEs introduce a new secure kernel/runtime (Figure~\ref{fig:risks}),
existing systems expect application developers to manually bridge this semantic gap~\cite{suciu2020horizontal}. The resulting ad-hoc approaches have exposed TEE-enabled applications to severe attacks across these interaction layers~\cite{suciu2020horizontal, khandaker2020coin,van2019tale,schwarz2019practical,weichbrodt2016asyncshock,weiser2019sgxjail,SchwarzMichael2017MGEU,machiry2017boomerang}. Even worse, mitigations based on sanitising the existing TEE interfaces are failing due to the wide attack surface~\cite{suciu2020horizontal,khandaker2020coin}.

We analysed 41 existing TEE-based open-source applications to classify the vulnerabilities from these attacks (\S\ref{appendix:attacks}).We identified the following classes of vulnerabilities across the host/enclave boundary that leads to sensitive data compromise:


\tpara{Insecure threading and procedure calls:} Attackers can launch malicious threads from within the host application process to exploit synchronisation vulnerabilities such as TOCTTOU~\cite{wei2005tocttou,weichbrodt2016asyncshock} or other types of concurrency attacks~\cite{khandaker2020coin} on enclave interfaces (Figure~\ref{fig:risks}~(1,2,4)). 
This greatly limits the secure use of enclaves within a multithreaded application; $56\%$ of TEE-enabled applications we analysed used multiple threads within the application or enclave.
\tpara{Memory vulnerabilities:} Attackers take advantage of inadequate address space isolation in each world to launch ROP attacks~\cite{lee2017hacking,biondo2018guard,machiry2017boomerang} for extracting cryptography keys or bypass remote attestation (Figure~\ref{fig:risks}~(2,3)). Additionally, insecure shared memory buffers are important attack vectors for extracting secrets or compromising RPC interfaces~\cite{khandaker2020coin}. 

\tpara{Insufficient privilege separation:}
A compromised or malicious third-party enclave can collect sensitive data~\cite{schwarz2019practical,marschalek2018wolf} and leak them using OS facilities such as files, network sockets or pipes (Figure~\ref{fig:risks}~5--10).
Attackers can launch horizontal privilege escalation (HPE) attacks~\cite{suciu2020horizontal} to compromise other process via a misbehaving enclave, or launch BOOMERANG attacks~\cite{machiry2017boomerang} to gain control of the host OS by tricking the secure world into modifying host kernel memory. No applications we studied implemented even a simple form of access control which would mitigate these vulnerabilities.

\input{attacks.tex}

Table~\ref{cvetab} summarises the attack vectors used by the TEE applications we studied. TEE applications use their OS interfaces to access system resources from userspace; e.g. threads, memory regions, IPCs or RPCs, files, and network sockets. Misusing these objects either directly permits the earlier attacks, or increases the attack's damage by propagating vulnerabilities or transferring extracted secrets to untrusted sources. 38 out of the 41 TEE-enabled applications we analysed depend on at least three of these facilities.
Therefore to comprehensively mitigate these classes of vulnerabilities, we need a system to selectively enable the protection and secure sharing of these objects across both the secure and normal worlds. 

We thus present Sirius, a framework for strongly compartmentalising and securely sharing systems resources and application data across normal world and enclave userspaces. It does so by extending the normal world and enclave kernels with interfaces that~\rom{1}~allow applications to compartmentalise sensitive data in both worlds;~\rom{2} enforce intra-process protection such that these compartments are protected even in multi-threaded applications with shared address space; and~\rom{3} support flexible and mutually distrustful access control across systems objects within both worlds.

Since TEE systems require running kernels on different privilege levels, we need a decentralized approach for enabling mutually distrustful compartments in each world. Sirius adopts decentralised information flow control (DIFC)~\cite{myers2000protecting} techniques for its unified access control to enforce mutual distrust, which provides application programmers with a simple-to-use labeling interface to control the flow of their private data.
Our implementation of Sirius on ARM Trustzone runs efficiently on commodity hardware (e.g. a Raspberry Pi). We have ported complex multithreaded use cases such as the Apache webserver, the Darknet ML framework, and safety-critical applications such as autonomous vehicles or medical devices that rely on secure data distribution services. Sirius framework provides a simple userspace API, and straightforward extensions to the Linux kernel and the TrustZone kernel.  In return, the applications are comprehensively protected from sensitive data leakage via the vulnerabilities described earlier and with small performance overheads on commodity hardware.

%% file: attacks.tex
\begin{table}[t]
\newcommand*\feature[1]{\ifcase#1 -\or\LEFTcircle\or\CIRCLE\fi}
\resizebox{\columnwidth}{!}{%
  \footnotesize
  \begin{tabular}{|c|l|l|l|l|l|l|l|}
    \hline
    &  \bf Category    &\bf Num    & \bf Thread & \bf Memory & \bf IPC/RPC & \bf Priv   \\
    \hline
    \parbox[t]{2mm}{\multirow{7}{*}{\rotatebox[origin=c]{90}{\bf  TEE-enabled Applications  }}}
    
    & Reference monitor \& Auditing & 8  &$\feature1$ &$\feature2$&   $\feature2$    & $\feature2$  \\
    & Web apps  &7&$\feature2$&$\feature2$& $\feature2$    & $\feature2$\\
    & Data analytics &5&$\feature2$&$\feature1$& $\feature2$ & $\feature2$ \\

    & Key management  &4& &$\feature1$&$\feature1$   &  $\feature1$ \\

    & Attestation &2&&$\feature2$& $\feature1$   &$\feature1$ \\
    & Databases &4&$\feature2$&$\feature1$& $\feature2$    & $\feature2$\\
    & SSL/TLS  &5&$\feature1$&$\feature2$& $\feature2$ &  $\feature2$ \\
    & Blockchain  &6&$\feature1$&$\feature2$&$\feature2$   & $\feature2$ \\
    
    & HPE ~\cite{suciu2020horizontal}  &  95 &&& $\feature2$  & $\feature2$ \\
     & BOOMERANG~\cite{machiry2017boomerang}  &  &&$\feature2$& $\feature2$ &$\feature1$ \\
    & COIN Attacks~\cite{khandaker2020coin}  &10&$\feature2$&$\feature2$&  $\feature2$    & $\feature1$ \\

    \hline
    
        \parbox[t]{2mm}{\multirow{4}{*}{\rotatebox[origin=c]{90}{\bf TCB Vulnerabilities }}}

                & \cve{2019-1010298}  &&$\feature1$&$\feature2$&  $\feature2$   & $\feature1$ \\

            & \cve{2018-11950}  &&$\feature1$&$\feature1$&  $\feature2$   & $\feature2$ \\

        & \cve{2017-8252}  &&$\feature1$&$\feature1$&  $\feature1$   & $\feature2$  \\
    & \cve{2017-8276}  &&$\feature1$&$\feature1$& $\feature2$  &$\feature2$  \\
    & \cve{2016-10297}  &&$\feature2$&$\feature1$&$\feature2$   & $\feature1$ \\
        & \cve{2016-5349} &&$\feature1$&$\feature2$&$\feature1$&$\feature2$   \\
    & \cve{2016-2431}  &&&$\feature2$& $\feature2$  & $\feature2$     \\
        & \cve{2015-4422}  &&&$\feature2$& $\feature1$      & $\feature2$\\

    \hline

  \end{tabular}}
  \begin{tablenotes}
\item $\feature2$ { }Object/feature is involved, $\feature1$ { }partially involved
\item $\text{{ }Priv:}$ Inadequate Privilege separation

\end{tablenotes}
  \caption{\label{t:cve-table} Summary of analyzing TEE-enabled applications and known attacks. Sirius fully addresses these threats or significantly weakens the damage. 
 }
  \label{cvetab}
\end{table}

%% file: back.tex
\section{Background \& Threat Model}\label{back}

Secure enclaves are a hardware facility that allow for partitioning sensitive data and associated computation away from the ``normal'' world.  Widespread TEE hardware includes ARM TrustZone (TZ) and Intel SGX, and they both support running a separate kernel within a secure world. Figure~\ref{fig:risks} illustrates the architecture of a typical (existing) TEE application on ARM TrustZone.

This paper focuses on the changes required to the host and ARM TZ-based enclaves to support Sirius' stronger security guarantees. We chose TZ for our prototype since its secure world is more powerful than other enclaves---and so exploited TZ enclaves can easily lead to a full OS compromise. TZ also does not directly support attestation as SGX does~\cite{intel2019linuxsgx}, which increases the possibility of hosting malicious enclaves. Also, billions of embedded devices use TZ-based TEEs, which requires Sirius to be resource-efficient. 
Porting Sirius to x86-64 requires straightforward engineering that we discuss later (\S\ref{diss}).  We will first describe how TZ hardware works (\S\ref{tzback}) and then elaborate on the threat model (\S\ref{sec:threats}).

\subsection{TrustZone Architecture}\label{tzback}
In ARM Cortex processors, security extensions or TrustZone is implemented by splitting each physical core into two virtual CPUs. Depending on the value of the Non-Secure (NS) bit, hardware resources (e.g., DRAM or peripherals) may run either in the secure world (SW) or the normal world (NW), where each run a separate software stack. TrustZone's one-way security model isolates SW by restricting NW to only its own resources; however, code running in the SW can access memory and I/O assigned for both worlds. 

Each world has its own user-mode (EL0/SEL0) and kernel-mode (EL1/SEL1). The control transition between the two worlds happens through a Secure Monitor Call (SMC) instruction that invokes the secure monitor code, which runs at the highest privilege level (EL3).
Although the TZ software stack and API interface security are not uniform across different devices, the widespread implementations (e.g., OPTEE, Kinibi, Teegris, QSEE) follow GlobalPlatform's ~\cite{DRTM} TEE specification. It requires enclaves or trusted applications (TAs) to run in SEL0 as processes with separate address spaces and communicate with TZ OS kernel, which runs in SEL1, via supervisor calls (SVCs). Usually, there are also privileged TAs like Trusted Drivers (TDs) in Kinibi~\cite{berard2018kinibi} or Pseudo TA (PTAs) in OPTEE~\cite{optee,tarkhani2019snape} that have access to a richer set of functionalities and SVCs to map physical memory, setting peripherals, use threads, and make SMC calls directly. OPTEE runs these privileged TAs directly as a TZ kernel driver in EL1. 

RPC requests between the two worlds consist of the TA identifier (e.g., UUID), a command ID that dictates which function to run, and a shared buffer for arguments or data transfer. TEE kernel driver in NW allocates shared memory from the host application's heap and only checks buffer sizes and direction flags as a basic security mechanism. Inadequate authorization easily allow communications between any set of applications and TAs. 
RPCs can be either stateful or stateless.  In stateless RPC, TA receives client application (CA) data, processes it, and returns a result without retaining any data across invocations. On the other hand, in stateful RPCs it persists some CA data across multiple invocations as a session state and global variables (.bss section) or inside persistent storage for larger objects. Typical secure storage in TZ is implemented by encrypting memory blocks that are stored in the NW's file system. For Example, OPTEE uses a secure storage key (SSK), enclave storage key (ESK), and file encryption key (FEK) for encrypting a persistent storage area in its boot file system. The per-device SSK is generated as a function of the unique hardware key and chip ID. The SSK must be stored in secure DRAM that is not accessible by the normal world, and will be used to derive the ESK.

\subsection{Threat Model}\label{sec:threats}

Sirius strengthens the original threat model of TEEs to cover the wide range of attacks originated from exploiting the vulnerability classes discussed earlier (\cref{intro}). Our Sirius threat model explicitly addresses attacks from compromised host applications, misbehaving enclaves, and various insecure interactions of threads running in both the secure and normal worlds (with each other and with shared systems resources).

We consider a commodity system running software from numerous independent vendors. Third-party library vendors and application developers may include enclaves to protect their secret data (e.g., cryptographic keys or intellectual property) or various security enforcement or auditing tasks. The Sirius compartmentalisation mechanism allows developers to control the flow of their data within kernel objects (e.g. file descriptors) in both worlds. 

Given the large number of vulnerabilities shown in Table~\ref{cvetab}, we assume a userspace attacker in both the secure and normal worlds, who could gain full control of a thread inside the host application or enclave and use OS services, memory operations and spawn more threads up to the resource limits. The attacker combines exploits to:

\begin{itemize}
\setlength\itemsep{0.1em}
\item interfere with other threads via crafted IPC/RPC requests, concurrent calls, shared memory access, or other systems resources~\cite{weichbrodt2016asyncshock,khandaker2020coin}.
\item exploit the TEE driver \texttt{ioctl} interface vulnerabilities (e.g., CVE-2015-4421 and CVE-2015-4422) or extract other thread's secrets or to gain full control of the host OS~\cite{machiry2017boomerang,lee2017hacking,biondo2018guard}.
\item bypass address space randomisation~\cite{seo2017sgx} by targeting the non-randomised runtime that handles transitions between the two worlds.
\item leak extracted secrets through untrusted threads and other system objects such as files, sockets, or pipes. 
\end{itemize}


The Sirius security model considers each thread to be a security principal and enables them to define a wide range of security policies based on mutual distrust. Sirius does not distinguish between a thread with root access or running inside a privileged enclave; it restricts unauthorized flow and access of private data as long as the security policies are specified and enforced correctly. 
This is why Sirius' security policy enforcement is decentralised between the host OS and the TEE kernel; vulnerability propagation and full-system compromise is far harder even if one of the kernels is compromised.  
Our work does not target microarchitectural covert or side-channel attacks ~\cite{xu2015controlled,van2018foreshadow,zhang2016truspy,chen2019sgxpectre,schwarz2019zombieload,lipp2018meltdown,kocher2018spectre}.

%% file: overview.tex
%

\section{Overview}\label{sit}

\subsection{Information Flow Model for TEEs}\label{sitee}
Since secure and normal worlds in TEE systems have their own security requirements and kernels, extending centralised security models such as MACs~\cite{setools}, system call filtering~\cite{seccomp}, or namespaces only allows static coarse-grained security policies. 
Prior work shows enabling DIFC allows complex applications and distributed systems to define a wide range of security policies and significantly improves their security~\cite{hunt2016ryoan,fernandes2016flowfence,zeldovich2008securing,nadkarni2016practical,roy2009laminar,krohn2007information,zeldovich2006making,myers2001jif}. Unlike classic IFC~\cite{denning1976lattice}, DIFC allows \textit{every security principal} to define trust boundaries via a set of labels drawn from a partially ordered set and allows communication if the labels satisfy an ordering. However, it is essential to provide developers with an understandable and practical programming model and implementation.
We extend DIFC principles to work with TEE systems via
\rom{1} low-overhead thread-granular enforcement of labels within kernel objects; 
\rom{2} strong isolation and secure sharing across multiple untrusting kernels on the same host, and \rom{3} secure label management and storage.

Our labeling model is inspired by Aeolus~\cite{cheng2012abstractions} and Flume~\cite{krohn2007information} and is based on three key concepts: principals, tags, and labels. Each thread is a principal representing an entity with security interests, and unique tags provide principals with a way to categorize their information. Labels are sets of tags and are used to control information flow. Privileges are represented in form of two capabilities $\theta^{+}$  and $\theta^{-}$ per tag $\theta$, that are stored in each thread's capability list $C_{t}=C_{t}^+ \cup C_{t}^-$  ($\theta^{+} \in C_{t}^+$ and $\theta^{-}  \in C_{t}^-$). These capabilities enable adding or removing tags to or from labels (similarly to Flume). Each thread $t$, has secrecy ($S_{t}$) label, which reflects confidentiality of information, and integrity ($I_{t}$) labels, which reflects the integrity of information, and a set $D_{t} \subseteq C_{t} $ that stores all tags for which $t$ has both privileges (full control). 
Thread labels are mutable: as a thread executes, its labels can change to reflect the secrecy and integrity of the information the thread has observed, subject to the rules defined below. Sirius maintains security state for each thread, consisting of its two labels, capabilities, and associated principal inside its handling kernel.

Information flow from a source $\alpha$ to a destination $\beta$ is allowed only if $S_{\alpha}\subseteq S_{\beta}$ to ensure that confidentiality is maintained as data propagates, and $I_{\beta}\subseteq I_{\alpha}$ to keeps track of influences of low-integrity entities. A thread can manipulate its labels by adding and removing tags, and all label manipulations must be done explicitly to avoid the label explosion problem~\cite{nadkarni2016practical}. Adding a tag to a secrecy label and removing a tag from an integrity label are safe operations since the thread only increases its contamination or reduces its integrity. However, declassification (removing a tag from a secrecy label) and endorsement (adding a tag to an integrity label) are unsafe operations and require the thread to be an owner or an authority~\cite{cheng2012abstractions}. When a thread creates a tag, it has authority for that tag. Subsequently, authority can be delegated to other threads to via \textit{grants} that can also be \textit{revoked}. Sirius maintains the delegation hierarchy for each tag, which form directed acyclic graph. Sirius allows transitive revocation that removes a particular link from the principal hierarchy or delegation graph only if the thread has the authority.


\begin{figure*}[t]
\centering
\includegraphics[width=\linewidth]{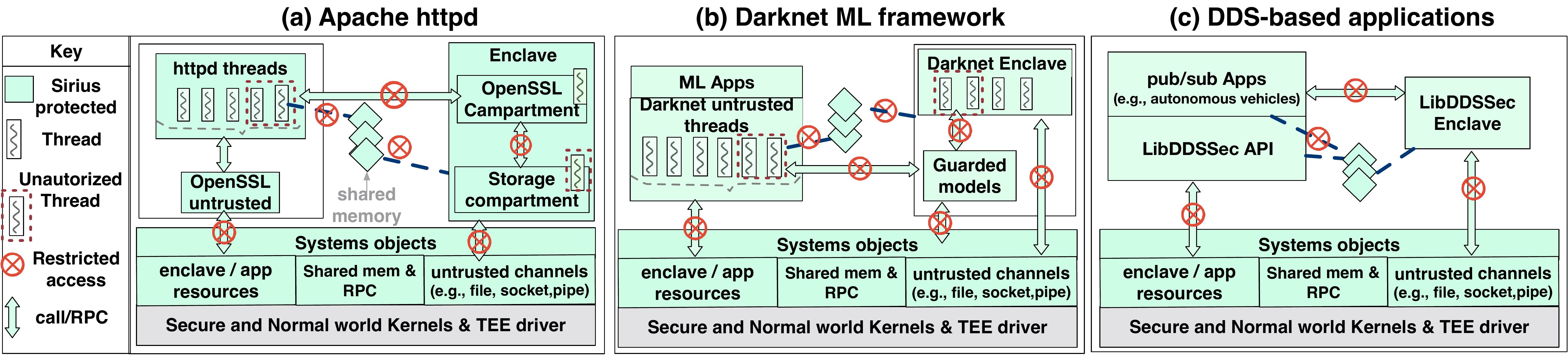}
\caption{High-level architecture of Sirius-protected applications }
\label{happs}
\end{figure*}


%% file: design.tex

\subsection{Sirius Design and API}\label{over}

Unlike previous TEE systems and frameworks, Sirius does not limit application partitioning to only coarse-grained trusted (i.e., in-enclave) and untrusted components (host application). Sirius allows further compartmentalisation inside the enclave and applications; each could have mutually distrustful compartments with different security policies. Note that in-enclave compartments are still part of an enclave; only now, Sirius isolates each compartment address space via new isolated memory compartment (IMC) abstraction(\S\ref{imc}),
as well as other required layers of interaction between these compartments. 
Before explaining Sirius' details, we briefly illustrate how to compartmentalise a TEE-enabled Apache webserver (in Figure~\ref{happs}a).

\tpara{Partitioning:} The developer uses Sirius APIs to define two in-enclaves compartments; a single-threaded OpenSSL compartment to run cryptographic operations and a storage compartment to store private keys and certificates. The developer then defines the enclave's interfaces via an API. The build-system cross-compiles the code into an executable enclave binary and generates UUID and per-enclave security keys, which are persistently stored by the TZ OS.  

\tpara{Compartmentalization:} The webserver needs to enforce mutual distrust between the enclave compartments and the normal world. The in-enclave OpenSSL compartment only needs access to the information required to establish a session key and no other user data. 
Only a subset of Webserver threads that handles OpenSSL operations are authorized to communicate with the enclave and access to the protected shared memory compartment between the webserver and enclave.  Sirius provides a malloc-style memory allocator that replaces traditional memory operations, such as in the OpenSSL EVP code. 
Sirius also labels the private keys files for the TLS negotiations to not be accessible to webserver threads. 

\tpara{Enforcement:} When the webserver starts, it creates a thread that can now launch the enclave and {\em only} that thread has the right label to interact with its compartments. The thread then grants the OpenSSL compartment direct RPC access to transfer secrets to the storage compartment and proceeds to revoke its own access to the secure storage. The normal world thread has now dropped its privileges and the storage compartment can only communicate with the OpenSSL compartment or its own labeled and encrypted filesystem.

\input{labels.tex}

The webserver uses a thread pool to handle incoming requests. The Sirius version can simply launch the same number of threads within the normal world (to handle external connections) and the OpenSSL enclave (to generate TLS session keys), and dynamically register memory regions so that the worker thread for a given connection only has access to its own session key. If a connection is compromised, that thread has no privileges to do anything beyond reading the one session key. If an enclave thread is compromised, it cannot leak its secrets to the outside world or access user data from the webserver.
 
We next explain the Sirius enclave lifecycle (\S\ref{senclave}) and our memory compartmentalization mechanisms (\S\ref{imc}). We do not describe the partitioning toolchain further, as it is largely consists of build system concerns.

\subsection{Sirius Enclave Lifecycle}\label{senclave}

Sirius modules extend normal world kernel (NK) and secure world kernels (SK) to enforce secure data flows.
An application thread $p$ calls \texttt{s\_create\_enclave} to spawn an enclave (see Table~\ref{scapi} for the interface). The NK creates a random secrecy-tag $x$ and adds it to the thread's secrecy label $S_{p}$ and ownership list $D_{p}$.
The NK then transfers the message to the secure monitor (SM) (i.e., a part of the SK codebase) via an SMC call (Figure~\ref{fig:labels}~\ding{172}a). The SM creates and persistently stores a new tag $y$ for the enclave and notifies the SK to assign both tags to a new enclave userspace thread $e$, by updating its empty labels to $S_{e}\{x,y\}$ and $D_{e}\{y\}$. The SM enforces message safety from $p$ to enclave $e$ by checking that
$S_{p} - D_{p}\subseteq S_{e} \cup D_{e} \land I_{e} - D_{e}\subseteq I_{p} \cup D_{p}$.
The SM passes the $y^{+}$ capability to the NK for updating $S_{p}$ to enable bidirectional calls (Figure~\ref{fig:labels}~\ding{172}b).

Both threads have each other's secrecy-tags but with only the \texttt{plus} capability. The SK is the only authority for declassifying (via \texttt{s\_declassify}) an enclave tag as well as all shared objects between the two worlds. Both kernels check RPC requests' safety between the two worlds, including function calls via \texttt{s\_ecall/ocall}, and ensures that no unauthorized thread can jump to an enclave entry. Both kernels drop unauthorized messages and kill the violating thread (Figure~\ref{fig:labels}~\ding{173}).

Each owner or authority thread (that has the right capabilities) can grant privileges to another thread via \texttt{s\_grant}, and revoke previously delegated privileges by calling \texttt{s\_revoke}.
The owner thread can also restrict any access or modifications of its object state by calling \texttt{s\_access\_disable}, which temporarily alters the object's tag until \texttt{s\_access\_enable} is called by the thread. This is particularly useful for fast intra-process access restriction when adapting untrusted code or libraries.
Child threads do not inherit labels by default (e.g., in the style of \texttt{fork}) as this makes it difficult to reason about security~\cite{baumann2019fork}. The parent thread can explicitly create a child with specified labels as an argument of \texttt{s\_clone}. 

Each NW thread can also use conventional Linux syscalls to access resources such as files, sockets, or pipes. We extend some syscalls such as \texttt{open} with extra flags (\texttt{SLABEL/ILABEL}) to create a labeled file, as are in \texttt{clone}, \texttt{create} and \texttt{pipe}.  Once labeled, Sirius controls the information flow within all operations on them via extended kernel abstractions (\ref{vfs}). The other substantial new feature in Sirius is the intra-address space memory compartmentalization, described next.

\subsection{Address Space Compartmentalisation}\label{imc}

Sirius's compartmentalization mechanism requires intra-address space memory protection that is not provided in existing OSs. We designed a new memory compartmentalisation abstraction to overcome this limitation efficiently. It introduces isolated memory compartments (IMCs) and enforces DIFC on these address space objects to isolate them across threads in the same process.

 \input{extra.tex}

Each IMC contains contiguous segments of virtual memory (VM).
Any thread can create one or more IMCs via \texttt{imc\_create} (An IMC can be kernel- or hardware- backed as explained in ~\cref{uhmm}), and its handling kernel assigns a new secrecy label to each. Hence, by mapping the same memory to different IMCs, Sirius can provide a separate memory view for associated threads. 
This enables different threads to have different privileges assigned for a shared memory block. When a thread has the $\theta^{+}$ capability for IMC $\theta$, it gains the privilege to access the IMC with the permission set by the IMC owner via \texttt{imc\_grant}. Having the declassification capability allows the thread to modify the IMC's memory layout by adding/removing pages via \texttt{imc\_mmap/munmap}, change IMC permissions via \texttt{imc\_mprotect}, or transfer the content to untrusted sources (e.g., copy to a file, or share with other thread). 

Shared IMCs between two worlds are always allocated in NW. 
Figure~\ref{fig:labels}~\ding{174} shows how both enclave $e$ and thread $p$ have access ($m1 \in S_{e} \land m1 \in S_{p}$) to the shared IMC $M_{1}$ and only with default permission set of $<p>$.
So enclave $e$ can not access $M_{2}$ and though thread $p$ has $m2\in S_{p}$ it is restricted for doing unautorzed operation $<q>$ on $M_{2}$. 
Since enclave is the owner ($m1 \in D_{e}$), $p$ can only access the IMC with default permissions and cannot modify the default state. The kernel restricts any unauthorized transfer of IMC's content via \texttt{memcpy}, \texttt{mmap}, or other unauthorized channels such as files, or pipes.
Both kernels protect IMCs against any unauthorized operations from both worlds' threads in SEL0 and EL0.
The code~\cref{code} shows how two threads can have IMCs with different privileges (\texttt{imc\_grant}) that enable a separate view of a shared memory (via \texttt{imc\_mmap}). \texttt{s\_clone} creates a labeled thread and maps IMCs into it.

\input{code.tex}

The next code snippet uses the IMC for memory protection within a single thread.
The application uses \texttt{imc\_malloc} call to allocate contiguous memory blocks within the IMC, \texttt{imc\_free} to deallocate memory, or \texttt{imc\_mprotect} to change its permissions. The thread has fine-grained control over its IMCs and can even protect them against untrusted code (e.g., unsafe third-party libraries) through the \texttt{s\_access\_enable/disable} calls. The SK checks all the operations of an IMC that are shared or owned by an enclave thread. The NK does the same for enclave-independent IMCs, so applications can shield their secrets even from their enclave. We later show how all these MCA features help to harden libraries such as OpenSSL with minimal in-enclave code ($96\%$ reduced enclave code) as an alternative to running it entirely inside an enclave (\cref{ahttpd}).

\input{code2.tex}



%% file: labels.tex
\begin{figure*}[t]
\centering
\includegraphics[width=\linewidth, height=3.9cm]{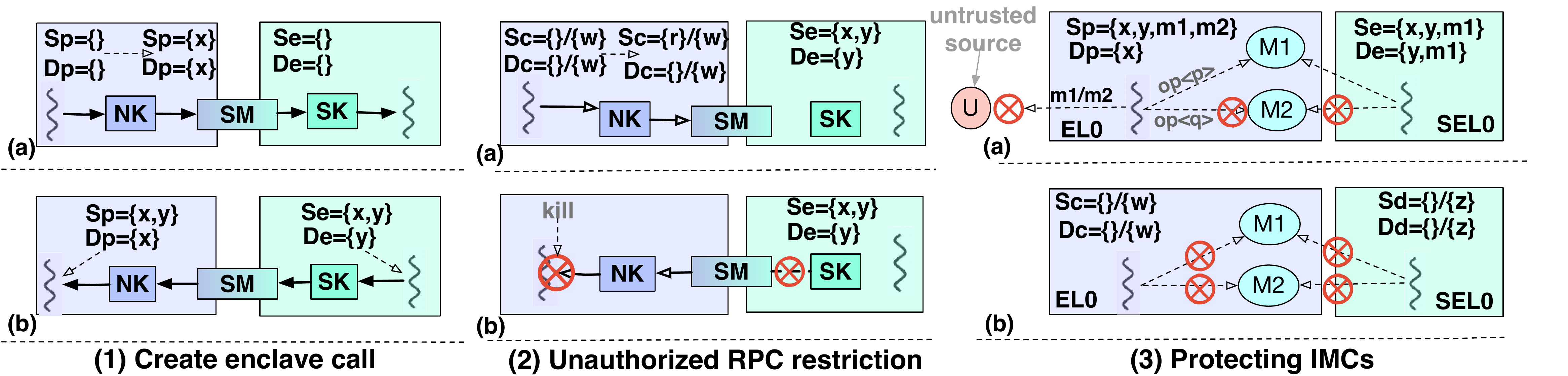}
\caption{Examples of information flow control in Sirius enclaves}
\label{fig:labels}
\end{figure*}

%% file: extra.tex
\begin{table*}[t]
\centering
\renewcommand{\arraystretch}{.8}
    \begin{tabular}{|ll|}
    \hline
    \textbf{Common enclave operations: }                                               & ~                                                    \\
    \texttt{s\_create\_enclave $\rightarrow$ eid}                                                  & create an enclave                                    \\
    \texttt{s\_ecall (rpc\_msg *, arguments)}                        & call from the host thread to the enclave             \\
      \texttt{s\_ocall (rpc\_msg *, arguments)}                      & call from the enclave to the host thread             \\
      \texttt{s\_delete\_enclave(eid)}                                      & delete an enclave                                    \\ \hline
      \textbf{Common security policy syscalls : }& ~                                                    \\
      \texttt{s\_grant/revoke (label\_info *, tid, dir)}                        & give/revoke thread privileges to objects           \\
      \texttt{s\_access\_enable/disable (label\_info *)}                                     & enable/disables access to set of objects                    \\
      \texttt{s\_clone (label\_info *,...) $\rightarrow$ tid}                                           & create a labeled child                               \\
      \texttt{open/socket/pipe(label\_info *,...) $\rightarrow$ ret}                                    & create a labeled kernel object                       \\ \hline
   \textbf{ API for Address space isolation:  }                                  & ~                                                    \\
      \texttt{imc\_create(hw-mode-flags) $\rightarrow$ imc\_id}                                            & create a new IMC (kernel- or hardware-backed)        \\
      \texttt{imc\_grant/revoke(imc\_id, P)}                                            & grants/revoke IMC the capability to imc with priv P        \\
      \texttt{imc\_malloc/free(imc\_id, ...)}                                         & allocate/deallocate memory within an imc                       \\
      \texttt{imc\_mprotect/mmap/munmap(imc\_id,...)}                                       & add/remove pages or change its  permission \\
      \texttt{imc\_kill(imc\_id)}                                                             & cleanup an IMC                                       \\ \hline
    \end{tabular}
    \caption{Summary of the Sirius application interface}
    \label{scapi}
\end{table*}

%% file: code.tex
\begin{lstlisting}[caption={Pseudocode of IMC API: threading example},captionpos=b,label={code}]
void imc_threading_test (void) {
//initialization ....
    imc_id[1] = imc_create(DEFAULT);
    imc_id[2] = imc_create(DEFAULT);
    // assign different permissions
    imc_grant(imc_id[1],MEMDOM_READ | MEMDOM_WRITE);
    imc_grant(imc_id[2], MEMDOM_READ );
    // map a same vm to IMCs
    imc_mmap(imc_id[1],addr,...,);
    imc_mmap(imc_id[2],addr,...,);
    // map imcs to threads
    tid1 = s_clone(imc_id[1],&fn ,..., SLABEL|flags);
    tid2 = s_clone(imc_id[2],..., SLABEL|flags);
//the rest....}
  
\end{lstlisting}

%% file: code2.tex
\begin{lstlisting}%[caption={Pseudocode of MCA API: example of IMC memory operations.},captionpos=b,label={code2}]
void imc_threading_test (void) {
//initialization ....
    imc_id = imc_create(DEFAULT);
    imc_grant(imc_id,
              MEMDOM_READ | MEMDOM_WRITE);
    /* allocate memory from imc */
    private_blk = (char*) imc_malloc(imc_id, len);
    /* make imc inaccessible */
    s_access_disable(imc,imc_id);
    //... untrusted computations ....//
    /* make imc accessible */
    s_access_enable(imc,imc_id);
    //... trusted computations ....//
    /* cleanup imc */
    imc_free(private_blk);
    imc_kill(imc_id);
//the rest....}  
\end{lstlisting}

%% file: implementation.tex

\section{Implementation}\label{imp}

The Sirius framework contains userspace API and kernel extensions at both secure and normal worlds to enable full-system isolation.
We have implemented Sirius on Linux kernel (\cref{linux}), which is the most used NW kernel for TEE systems, and we extend OP-TEE OS (\cref{stee}), which is a popular open-source TZ kernel.

\subsection{Linux modifications}\label{linux}
Sirius adds a new security module in the Linux kernel (version $4.19.42$) to govern information flows through fundamental systems objects (\cref{lsm}). It extends virtual memory abstractions to enable kernel- and hardware-backed intra-process isolation (\S\ref{mca}, \S\ref{uhmm}), with small additional modifications for enforcing DIFC within traditional kernel objects (\cref{vfs}) like files, sockets, and pipes.


\subsubsection{Sirius security module}\label{lsm}

Our designed-from-scratch and extremely lightweight Linux security module (LSM) implement one set of clear rules for enforcing DIFC within any primary kernel objects such as \texttt{inodes, tasks, IMCs}. It provides only necessary security hooks (e.g., \texttt{change\_label}, \texttt{check\_flow\_allowed}) that are used in the rest of the kernel to govern the information flow control. The LSM initialises required data structures such as the label registry that caches labels and capability lists per threads. We implemented a hash table-based registry to make operations (store/set/get/remove) on these data structures more efficient. The LSM also handles synchronisation for labeling operations using mutexes and atomic operations.

The LSM stores labels and metadata required for enforcing DIFC in each thread's \texttt{cred} structure. We modified \texttt{copy\_creds}  and \texttt{copy\_process}
to disallow \texttt{cred} inheritance by allocating an empty \texttt{cred} per thread. Some LSM's FS-specific hooks are used for managing \texttt{inodes} labels and enforcing the safe flow within files and directories; e.g. via the \texttt{inode\_permission} and \texttt{file\_permission} security hooks. The LSM provides similar hooks for DIFC enforcement within sockets and pipes, and IMCs. Sirius's LSM in total only needs 15 hooks that is significantly less than other LSMs such as SELinux (i.e., more than 200 hooks). 

\subsubsection{IMC implementation}\label{mca}

We extend the kernel VM layer to support efficient intra-address space isolation via IMC abstraction. Each IMC maintains a secrecy label and has a private virtual page table (\texttt{pgd\_t}) that is loaded into the \texttt{TTBR} register when the thread needs to do memory operations inside an IMC during a lightweight context switch. An internal IMC data structure maintains its address space range and permissions (Appendix A:\cref{imcstruct}).

Threads (or Linux tasks) in a process share the same \texttt{mm\_struct} that describes the process address space. Having separate \texttt{mm\_struct} for threads would significantly impact system performance, as all the memory operations related to page tables should maintain strict consistency~\cite{hsu2016enforcing}. Instead, we extend \texttt{mm\_struct} to embed IMC metadata within it as lightweight protected regions in the same address space (Appendix A:\cref{mmstruct}).
It stores a per-IMC \texttt{pgd\_t} for threads and other metadata for memory management, fault handling, and synchronization.

The standard Linux kernel avoids reloading page tables during a context switch if two tasks belong to the same process. We modified \texttt{check\_and\_switch\_context}
to reload IMC page tables and flush related TLB entries if one of the switching threads owns an IMC.
We further mitigate the flushing overhead using tagged TLB features and ARM memory domains (\cref{uhmm}).
We modify \texttt{mmap.c} to keep track of IMC-mapped memory ranges and add \texttt{imc\_mmap/mumap} operations.

The kernel \texttt{handle\_mm\_fault} handler is also extended to specially manage page faults in IMC regions, so an IMC privilege violation results in the handler killing the violating thread.

\subsubsection{Hardware-backed IMC}\label{uhmm}
We provide an \texttt{optional optimization} for IMC implementation by utilizing ARM memory domains (MDs)~\cite{arm2012architecture} if supported by the hardware. ARM-MDs are a lesser-known memory access control mechanism that is independent of paging. Each page table first-level entry has 4 bits allocated to support 16 memory domains. Access control for each domain is handled by setting a domain access control register (DACR) in \texttt{CP15}, which is a 32-bit privileged register. Changing domain permissions are low cost and do not require TLB flushes, and any access violation causes a domain fault.
The table below shows the four possible access rights for a domain.  

\input{domains.tex}

The optimised abstraction maps IMC to hardware domains instead of separate virtual page tables, and so supports up to 16 1MB-aligned IMCs. Sirius provide a separate set of kernel memory management functions similar to their Linux equivalents (e.g., \texttt{do\_mmap}, \texttt{do\_munmap} and \texttt{do\_mprotect}) for mapping IMCs to hardware domains.
Due to the reduced number of TLB flushes and faster context switches, using ARM-MDs improves the cost of Sirius threading by $38\%$ (\cref{mbenches}).
The hardware-backed IMC improves the performance of \texttt{imc\_mmap/munmap} by $48\%$ due to the simpler mechanism of memory mapping to the memory domain instead of virtual page tables.
It also improves the performance of \texttt{imc\_mprotect} ($1.14$x faster than \texttt{mprotect}) if the requested permission change matches one of the supported hardware options; otherwise, the cost is the same as \texttt{imc\_mprotect}. The optimised IMC also has a more lightweight fault handler that utilise domain faults instead of full page faults.

\subsubsection{Tracking flows within the kernel}\label{vfs}

We modified the VFS layer to enforce thread's security policies within all operations on \texttt{inode}, \texttt{file}, and VFS address space objects; these kernel abstractions are used to perform operations on unopened files and file handles (including sockets and pipes).
Most \texttt{inode} operations (e.g., \texttt{create}, \texttt{link}, \texttt{mknod}, \texttt{mkdir}, \texttt{permission}) require a lookup to find related \texttt{inodes} and \texttt{dcaches}; hence, we modified the kernel \texttt{namei} to disallow unauthorized information flow at early lookup stages.

We extended the \texttt{open} syscall with two new flags (\texttt{SLABEL} and \texttt{ILABEL}) that a thread can use to create a labelled file (e.g. \texttt{O\_CREAT} | \texttt{SLABEL}) and added \texttt{file/inode\_permission} security hooks on necessary places to disallow unauthorised file operations like \texttt{read}/\texttt{write}/\texttt{stat}/\texttt{seek}. A malicious thread may also try to map a labeled file to an address space object via \texttt{writepage}. Sirius checks that labeled files are only be mapped to IMCs with the right labels via \texttt{imc\_mmap}.

The label of an \texttt{inode} protects its contents and its metadata. In a typical filesystem tree, secrecy increases from the root to the leaves. To ensure writing a new entry in a parent directory does not disclose secret information, we disallow a thread with secrecy label $S\{x\}$ from creating a file with the same secrecy label in an unlabelled directory since it leaks information through the filename. Our LSM lets a thread with non-empty labels $S_{p}, I_{p}$ create a labeled file or directory with labels $S_{d}, I_{d} $, if the label change is safe and the thread can write to the parent directory with its current label.
Sirius stores normal files' labels in the extended attributes or in the secure storage if the file is an enclave-shared/owned object.

We also modified the kernel to enforce DIFC in socket operations like \texttt{create}, \texttt{listen}, \texttt{connect}, \texttt{sendmsg}, and \texttt{recvmsg}. This was done by placing security hooks in those functions and at the end of the lookup process (e.g., \texttt{sockfd\_lookup\_light}). All operations for unlabelled threads and unlabelled objects follow the traditional Linux access control mechanisms, so applications that do not use Sirius do not require any modifications.  Similarly, Sirius controls information flow within pipes, so a thread may read or write to a pipe as long as its labels are compatible. Sirius does not allow a labeled thread to \texttt{connect} to a socket unless that thread has the declassification capability for the accessed secrets. Messages that cannot be delivered are rejected silently to avoid leaking information by returning errors.


\input{tfs.tex}

\subsection{Secure world modifications}\label{stee}

We extend OP-TEE V3.4 secure kernel and monitor (optee\_os), and the TEE driver (\texttt{libTEEC}) to enforce DIFC within enclave threads, RPC messages, and IMCs. The original OP-TEE security model is based on GlobalPlatform~\cite{gplatform} API security checks. It checks RPC messages by validating arguments, buffer sizes, and directions flags
at every layer of privilege (\texttt{EL0}, \texttt{EL1}, \texttt{EL3}, \texttt{SEL1}, \texttt{SEL0}). However, these checks have been bypassed many times~\cite{linaro}. For shared memory, OP-TEE checks the address range, cache attributes, and size of allocated memory chunks. This is also insufficient in many cases (\cref{sec:threats}); for example, to avoid BOOMERANG attacks, the authors extended OPTEE with the CSR-based pointer verification~\cite{machiry2017boomerang}. Sirius's security model is based on fine-grained compartmentalisation and isolation rather than error-prone security checks.

\tpara{Security Monitor:} We first modified the optee\_os security monitor (\texttt{/core/sm}), which is the entry point of RPC messages between the two worlds and runs at the highest privilege level. We extended the monitor to label each enclave, and store labels and capability lists in \texttt{sm\_cred}, a new data structure. IFC over RPC requests is enforced by adding a security module similar to our LSM (\cref{lsm}) to the secure kernel. When an RPC is safe and leads to label changes, the monitor transfers its \texttt{sm\_cred} data structure to the secure and normal worlds to each update their thread labels accordingly. 

\tpara{Secure Kernel:} The unmodified OP-TEE secure kernel assigns a static number of threads for each enclave (\texttt{CFG\_NUM\_THREADS}). Execution of enclave threads is tied to the execution of the caller thread and scheduled by the Linux kernel.
The secure kernel uses several L1 translation tables (one spanning $4GB$) and some smaller tables spanning $32 MB$. 
The large translation table handles secure kernel mappings (\texttt{TTBR1}), and the small tables are assigned per thread and map enclave contexts to its dedicated VM. We also extend the secure kernel with IMC metadata and virtual page tables to enforce DIFC within enclave threads and address space objects similar to the normal world abstractions.

\tpara{Enclave Userspace:} 
We replaced the OP-TEE shared memory mechanism with an IMC-assisted one via new \texttt{ioctl} calls to the OP-TEE driver (e.g., 
\texttt{TEE\_IOC\_SHM\_SIRIUS\_ALLOC}). We also added support for the enclave-side versions of the IMC API. Enclave threads now benefit from Sirius's fine-grained compartmentalization and protected shared memory as described in \S\ref{mca}.

The original OPTEE supports a limited encrypted storage mechanism using a (non-POSIX) interface to the Linux filesystems.
While useful for storing enclave-related keys, it is impractical for applications with moderate I/O requirements (\S\ref{mbenches}).
We extended OPTEE to provide labeled access to the Linux FSs, allowing enclave threads to control their files without high overhead.

%% file: domains.tex
\begin{table}[htp]
\resizebox{\columnwidth}{!}{%

    \begin{tabular}{|l|l|l|}
    
    \hline
    Mode & Bits & Description                                                                     \\ \hline
    No Access   & 00 &  Any access causes a domain fault.                          \\\hline
    Manager     & 11 & Full accesses with no permissions check.\\\hline
    Client      & 01 &  Accesses are checked against page tables   \\\hline
    Reserved    & 10 &   Unknown behaviour. \\ \hline
    \end{tabular}}
\end{table}

%% file: tfs.tex
 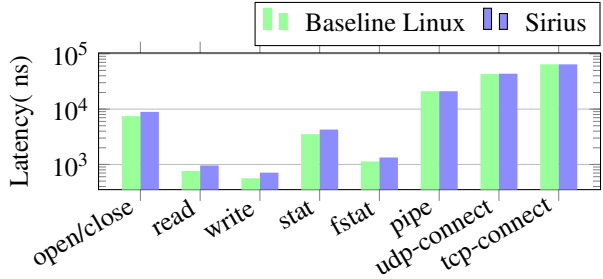
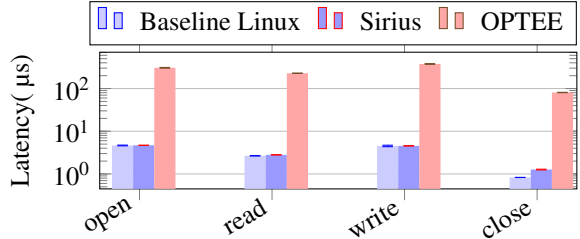
\begin{figure*}[ht]

\begin{subfigure}{.5\textwidth}

 \begin{tikzpicture}
    \begin{axis}[
        width=\textwidth,
        height=3.4cm,
        bar width=7pt,
        symbolic x coords={open/close,read,write,stat,fstat, pipe, udp-connect, tcp-connect},
        xlabel={},
        xtick=data,
        x tick label style={rotate=30,anchor=east},
        ybar=0pt,
        ymajorgrids=true,
        ylabel={Latency( ns)},
        ymode=log,
         legend cell align=left,
        legend style={
            at={(1,1.05)},
            anchor=south east,
            legend columns=2,
            column sep=1ex
        },
    ]
        \begin{scope}[
            draw=none,
        ]

        \addplot [draw=none, fill=blue!70 , green!40] coordinates {
        (read,765.7)
        (write,563.4)
         (stat,3524.7) 
        (fstat,1133.5)
        (open/close,7465.1)
        (pipe,21042)
         (udp-connect,43293)
        (tcp-connect,64396)};
         \addplot [draw=none, fill=blue!45] coordinates {
        (read,959.5)
        (write,711.1)
         (stat,4272.4) 
        (fstat,1336.1)
        (open/close,8911.9)
        (pipe,21053) 
        (udp-connect,43639)
        (tcp-connect,64401)};

        \end{scope}

\legend{Baseline Linux, Sirius }
    \end{axis}
\end{tikzpicture}
\caption{Common syscalls using LMbench.}
\label{lmbench}

\end{subfigure}\hfill
 \begin{subfigure}{.48\textwidth}
 \begin{tikzpicture}
    \begin{axis}[
        width=\textwidth,
        height=3.4cm,
        bar width=8pt,
        symbolic x coords={open,read,write,close},
        xlabel={},
        xtick=data,
        x tick label style={rotate=30,anchor=east},
        ybar=0pt,
        ymajorgrids=true,
        ylabel={{Latency( \textmu s)}},
        ymode=log,
         legend cell align=left,
        legend style={
            at={(1,1.05)},
            anchor=south east,
            legend columns=3,
            column sep=1ex
        },
    ]
        \begin{scope}[
            xshift={0.1*\pgfplotbarwidth},
            draw=none,
        ]

\addplot+ [draw=none, fill=blue!20,
            error bars/.cd,
                y dir=both,
                y explicit relative,
        ] coordinates {
          (open,4.644) +- (0,0.02)
            (read,2.662) +- (0,0.01)
            (write,4.527) +- (0,0.04)
            (close,0.828) +- (0,0.003)
        };

\addplot+ [draw=none, fill=blue!40,
            error bars/.cd,
                y dir=both,
                y explicit relative,
        ] coordinates {
            (open,4.679) +- (0,0.01)
            (read,2.802) +- (0,0.008)
            (write,4.549) +- (0,0.01)
            (close,1.268) +- (0,0.02)

        };

\addplot+ [draw=none, fill=red!35,
            error bars/.cd,
                y dir=both,
                y explicit relative,
        ] coordinates {
    (open,305.682) +- (0,0.02)
            (read,227.858) +- (0,0.01)
            (write,375.816) +- (0,0.02)
            (close,80.717) +- (0,0.01)
        };
        
        \end{scope}

\legend{Baseline Linux, Sirius ,OPTEE}
    \end{axis}
\end{tikzpicture}

\caption{Sirius's file protection vs OP-TEE secure storage.}

\label{sfs}

\end{subfigure}
\label{microbench}
\caption{Overhead of Sirius-protected kernel objects }
 \end{figure*}
 

%% file: eval.tex

\section{Evaluation}\label{eval}

We have so far explained how Sirius implements system-wide isolation to guard applications partitioned across normal and secure worlds (\S\ref{imp}). Sirius reduces the overhead of DIFC significantly by: \rom{1} enforcing and tracking labels in the kernel abstractions rather than userspace; and \rom{2} adding a new abstraction for address space compartmentalisation to achieve intra-process memory isolation (\S\ref{uhmm}).  We next examine the impact of these choices, with microbenchmarks (\S\ref{mbenches}) and porting applications (\S\ref{sec:apps}).

Our evaluation is done on Raspberry Pi 3 Model B~\cite{rpi3} with a 1.2 GHz 64-bit quad-core ARM Cortex-A53 processor with 32KB L1 and 512KB L2 cache memory, running a 32-bit unmodified Linux kernel version 4.19.42 and glibc 2.28 as the baseline.

We modified Linux to enforce DIFC within the normal world systems objects (\S\ref{linux}). Our kernel patch only adds $\approx10K$ LoC, of which the Sirius's standalone LSM (\cref{lsm}) is $\approx 5.4K$ new LoC and the IMC modifies $\approx 2.5K$ LoC within the virtual memory layer (\cref{mca}). The hardware-backed IMC required fewer changes as it bypassed much of the existing Linux code by using hardware domains (\cref{uhmm}). The remaining changes are mostly done to VFS and networking layers (\cref{vfs}).
We extended OPTEE V3.4 to enforce DIFC within the secure world systems objects. Our modifications add $\approx 2K$ LoC to the security kernel and monitor, and $\approx 3K$ LoC to the TEE driver and userspace API. 

\subsection{Microbenchmarks}\label{mbenches}

{\em What is the overhead of Sirius on a baseline Linux kernel? How much does the Sirius LSM affect the performance of general OS services such as filesystem, networking, threading, and memory operations? How effective is the use of hardware memory domains for optimizing IMCs?}

\tpara{Linux:}
We used LMbench 3.0~\cite{mcvoy1996lmbench} to evaluate the overall overhead of our Linux modifications compared to the baseline kernel (Figure~\ref{lmbench}). The results show that enabling Sirius on all file systems causes $\approx 1.2$x slowdown. Figure~\ref{sfs} shows that Sirius protection is $\approx 81$x faster than the OPTEE secure storage mechanism, which uses a heavyweight forwarding mechanism to the Linux. The Sirius labelling approach has reasonable overhead and makes it far easier to securely share systems resources across the host application and enclave. 
Latency overhead is $\approx 0.7\%$ on LMbench networking benchmarks.

\tpara{Threading:} We tested the cost of creating and joining (using \texttt{waitpid}) Sirius threads using \texttt{clone} with the new \texttt{SLABEL} flag that creates a secrecy-tagged thread.  
We also run \texttt{pthread} and \texttt{fork} microbenchmarks on the baseline kernel.
The table below shows the average latency (\textmu s) of $100000$ runs with 1MB and 2MB heap sizes.

\input{thrd.tex}

Forking is far more expensive than baseline threads with shared address space. The Sirius threads are slower than \texttt{pthreads} due to the overhead of our IMC-based memory isolation, but with our hardware-backed IMC optimization, Sirius threads add only $2.5\%$ overhead compare to \texttt{pthreads}. This highlights the importance of utilizing HW-based VM tagging for optimizing IMC.


\tpara{Enclave operations:}

Our changes to OP-TEE replaced checks spread throughout it with Sirius-enabled enclave operations that improves security and performance. The table below reports the average of $20000$ runs of our microbenchmark that shows Sirius secure world is $\approx 8.3\%$ faster than unmodified OPTEE with baseline Linux.

\input{optee.tex}

\tpara{Memory allocation:}
We next test our memory compartmentalisation overhead, first for shared memory allocation. We test baseline OP-TEE, and the BOOMERANG~\cite{machiry2017boomerang} CSR code that adds additional pointer verification to OP-TEE, with our IMC-based approach. The following results show that Sirius shared memory protection outperforms both by $\approx 16\%$ and $\approx 31\%.$ respectively, while providing stronger and thread-granularity address space isolation.

\input{microbench.tex}


\tpara{Memory protection:}

We measure the cost of memory protection for baseline Linux where protection is per-process, and on Sirius threads where protection is per-thread and either implemented in software (\S\ref{mca}) or hardware (\S\ref{uhmm}).
The next graph shows the average results of $10000$ runs of our microbenchmark comparing the cost of \texttt{imc\_mprotect} with \texttt{mprotect} on baseline kernel. The results show \texttt{imc\_mprotect} is $1.12$x slower than \texttt{mprotect}, but the hardware-backed Sirius \texttt{imc\_mprotect} is $1.14$x faster than baseline for some permissions (none and r/w) that supported by \texttt{DACR} register and do not need a TLB flush (\S\ref{uhmm}). 

\input{optmized.tex}

%
%

\input{openssl.tex}

 \subsection{Protecting Applications with Sirius}
 \label{sec:apps}
 
Sirius aims to make the usage of TEE systems more widespread in conventional applications, as well as improve the security of existing TEE-assisted applications. We chose three applications to comprehensively adapt to the Sirius APIs.  Firstly, the popular Apache httpd can be adapted to run with reasonable overhead under Sirius (\S\ref{ahttpd}).  Then we port two popular TEE-assisted applications -- a machine learning framework (\S\ref{dnet}) and DDS-based control system (\S\ref{dds}), and show how our system-wide isolation improves their security and performance at the same time. Figure~\ref{happs} illustrates the ported architecture of all three applications.

%
%
%


 \subsubsection{Apache httpd and OpenSSL}\label{ahttpd}

We earlier described the high-level architecture of the enclave-protected httpd in Sirius (\S\ref{over}).
We built a TEE-assisted OpenSSL using two enclave's compartments and only modified $\approx 2.4K$ LoC out of $\approx 533K$ LoC. The ported httpd protects all private keys, session keys, and certificates and operations on them from any unauthorized thread by defining trust boundaries in both normal and enclave worlds. 
It forbids a malicious enclave thread from transferring secrets through uncontrolled channels to another enclave, or via untrusted memory, or via a file or networking sockets.
A malicious httpd worker thread cannot compromise the enclave by crafting RPC requests or modifying shared memory or even by gaining root privilege\footnote{See CVE-2019-0211 or CVE-2019-0217, among others.} unless also compromising the host kernel and security kernel to obtain the right labels. It also provides in-depth mutually distrustful isolation of stored data, metadata, and binaries on the host filesystem for both enclaves and httpd.

We modified OpenSSL \texttt{libcrypto} to support a protected heap via a shared IMC owned by our EVP\_enclave. All the data structures that store private keys (\texttt{EVP\_PKEY}) now use the Sirius IMC memory operations such as \texttt{imc\_malloc/free} that is replaced with original \texttt{CRYPTO\_malloc/free}. The EVP\_enclave thread is the owner of this protected heap. Sirius protects the secrets that are being processed in this memory region, usually via cryptography operations such as \texttt{EVP\_Encrypt/DecryptUpdate} or \texttt{pkey\_rsa\_encrypt/decrypt}. The main httpd thread grants the \texttt{plus} capability to the EVP\_enclave for communication with the enclave\_storage\_compartment to store encrypted content, keys, and certificates inside storage that is labeled to be hidden from other threads. The
 EVP\_enclave thread is also the owner of all the OpenSSL files and directories (e.g., \texttt{OPENSSLDIR}) to restrict unauthorised or accidental information leaks. 

Figure~\ref{fig:shttpd} shows the overhead of ApacheBench applied against the original OpenSSL library on a baseline kernel and the Sirius-assisted httpd. ApacheBench ran with a timeline of 5 minutes for each request size, with the TLS1.2 \texttt{DHE-RSA-AES256-GCM-SHA384} algorithm cipher suite. The results show that Sirius-enabled httpd adds $\approx 10.8\%$ overhead on multithreading benchmark. This is a very reasonable overhead for an application that now gains fine-grained isolation with defense-in-depth layers to protect its secrets against threats from both the normal and secure world, which was not possible without Sirius.

\subsubsection{Privacy-preserving ML}\label{dnet}

TEE-assisted ML frameworks such as DarkneTZ~\cite{mo2020darknetz} are designed to avoid membership inference attacks (MIA)~\cite{shokri2017membership} against ML models and training data~\cite{hunt2018chiron,salem2018ml,vannostrand2019confidential}. We modified DarkneTZ to protect it against attacks that require even finer-grained compartmentalisation.  

Darknet is a heavily multithreaded application that launches many threads for training and processing sensitive data that could potentially misbehave. Sirius ensures that only authorised threads can issue queries to the enclave, providing another layer of protection against MIA attacks. Sirius labels the ML models stored in the host filesystem to be hidden from any untrusted thread and restricts enclaves from transferring the models or any processing data to untrusted sources.

We ported OPTEE-based DarkneTZ to Sirius with only minor modifications (318 LoC) to provide full-system security guarantees. We modified the Darknet classifier (\texttt{classifier.c}) to launch secrecy-tagged threads for communicating with enclave layers and used regular threads for the rest of the data loading logic. We utilized Sirius-guarded RPC and IMC-based shared memory operations, and protected all sensitive resources such as configs (\texttt{/cfg}), models (\texttt{/models}), and data (\texttt{/data}) on the host OS. 

We evaluated the performance using AlexNet, which has five convolutional layers. We train a model with four layers outside and one layer inside an enclave using CIFAR-100~\cite{cifar} for both training and inference. 
CIFAR-100 includes 50k training and 10k test images belonging to 100 classes. The table below shows the Sirius overhead compared to OP-TEE and the baseline when all layers run in the normal world.

\input{ml.tex}

Sirius outperforms OPTEE-based DarkneTZ by $\approx 5.6\%$ and on average adds $\approx 9.2\%$ overhead compare to baseline Darknet, despite the improved layers of isolation and amount of data flowing across the normal and secure worlds.

\subsubsection{Secure DDS}\label{dds}

Security-sensitive IoT applications such as autonomous vehicles or medical devices use TEE-assisted data delivery service (e.g., ARM LibDDSSec) enclaves for security-critical tasks. 
Sirius hardens LibDDSSec handlers for authentication, protecting data samples, secret sharing, and certificate operations, which all require secure interactions with the normal world.
The changes ensure that all shared data from other nodes are protected while being processed (via IMCs) and while at rest (via labeled files). Only trusted threads can exchange safety-sensitive messages, control messages, critical system data (e.g. emergency start/stop), and sensor data (e.g. temperature, laser, camera).

\input{dds.tex}

We modified \texttt{dsec\_ca.c} to replace the OPTEE RPC and shared memory with Sirius-protected operations. Sirius restricts any node from leaking private content through uncontrolled channels by labelling associated sockets and files.
 We evaluate the overhead using the OPTEE-enabled LibDDSSec benchmarks that show Sirius improves the overhead by $0.05\%$.

%% file: thrd.tex
\begin{table}[htp]
\resizebox{\columnwidth}{!}{%

    \begin{tabular}{|l|c|c|c|c|}
    \hline
    Operation       &       {fork} & {pthread} & {s\_clone} & {hw s\_clone}    \\ \hline
    Launch (1MB)        & 280.24 & 31.17 & 51.80 & 31.98  \\ \hline
    Join (1MB)            &  832.45 &  1.10  & 3.78 &  1.70  \\ \hline
    Launch (2MB)         &  331.40 & 31.51 & 51.85 & 32.1 \\ \hline
    Join (2MB)         &  1126.69& 1.13 & 3.82(3) & 1.78 \\ \hline

    \end{tabular}}
\end{table}

%% file: optee.tex
\begin{table}[htp]
\sisetup{separate-uncertainty}
 \resizebox{\columnwidth}{!}{%
\begin{tabular}
  {@{} 
    l
    S[table-format = 3.2(2)]
    S[table-format = 3.2(2)]
    S[table-format = 3.2(2)]
  @{}}
  \toprule
    & \multicolumn{2}{c}{Latency ({\textmu s})} \\
    &  \multicolumn{1}{c}{OP-TEE} & {Sirius SK}  \\
  \midrule
      create enclave          & 99.82(2)            &     93.95(1)   \\
    delete enclave         &    30.02(1)   &      30.10(1)  \\
     enclave calls (ecall ocall)         &  22.68 (1)     &    20.14(3) \\

  \bottomrule

\end{tabular}}

\end{table}

%% file: microbench.tex
\begin{table}[htp]
\resizebox{\columnwidth}{!}{%
 \begin{tikzpicture}
    \begin{axis}[
        width=.8\textwidth,
        height=5cm,
        bar width=7pt,
        symbolic x coords={1,2,4,8,16,32,64,128,256,512},
        xlabel={\Large{Allocated Memory (KB)}},
        xtick=data,
        ymin=0,
         ymax=15,
        ybar=0pt,
        ymajorgrids=true,
        ylabel={\Large{Latency(\textmu s)}},
        legend cell align=left,
        legend style={
            at={(1,1.05)},
            anchor=south east,
            legend columns=4,
            column sep=1ex
        },
    ]
        \begin{scope}[
            xshift={0.1*\pgfplotbarwidth},
            draw=none,
        ]

\addplot+ [draw=none, fill=red!20,
            error bars/.cd,
                y dir=both,
                y explicit relative,
        ] coordinates {
            (1,8.087) +- (0,0.01)
            (2,8.186) +- (0,0.07)
            (4,8.682) +- (0,0.07)
            (8,9.581) +- (0,0.008)
            (16,9.630) +- (0,0.009)
            (32,10.069) +- (0,0.03)
            (64,10.933) +- (0,0.04)
            (128,12.163)  +- (0,0.09)
            (256,12.52) +- (0,0.08)
            (512,12.65) +- (0,0.09)
        };

\addplot+ [draw=none, fill=red!50,
            error bars/.cd,
                y dir=both,
                y explicit relative,
        ] coordinates {
            (1,10.1) +- (0,0.03)
            (2,10.16) +- (0,0.02)
            (4,10.8) +- (0,0.03)
            (8,10.8) +- (0,0.002)
            (16,10.9) +- (0,0.001)
            (32,12.5) +- (0,0.03)
            (64,12.5) +- (0,0.02)
            (128,14.3)  +- (0,0.01)
            (256,14.52) +- (0,0.02)
            (512,14.43) +- (0,0.01)
        };
        
\addplot+ [draw=none, fill=blue!40,
            error bars/.cd,
                y dir=both,
                y explicit relative,
        ] coordinates {
            (1,7.327) +- (0,0.07)
            (2,7.278) +- (0,0.005)
            (4,7.094) +- (0,0.03)
            (8,7.166) +- (0,0.1)
            (16,7.66) +- (0,0.1)
            (32,9.42) +- (0,0.1)
            (64,10.61) +- (0,0.08)
            (128,10.18)  +- (0,0.09)
            (256,10.47) +- (0,0.09)
            (512,10.43) +- (0,0.09)
        };

        \end{scope}

\legend{OPTEE,OPTEE-CSR,Sirius}
    \end{axis}
\end{tikzpicture}}

\end{table}

%% file: optmized.tex
\begin{figure}[htp]
\resizebox{\columnwidth}{!}{%
 \begin{tikzpicture}
    \begin{axis}[
        width=0.6\textwidth,
        height=4cm,
        bar width=7pt,
        symbolic x coords={RO, WO, EO, R/W, NO-ACC},
        xtick=data,
        ymin=0,
         ymax=3.5,
        ybar=0pt,
        ymajorgrids=true,
        ylabel={Latency(\textmu s)},
        legend cell align=left,
        legend style={
            at={(1,1.05)},
            anchor=south east,
            legend columns=3,
            column sep=1ex
        },
    ]
        \begin{scope}[
            xshift={0.1*\pgfplotbarwidth},
            draw=none,
        ]

        \addplot+ [draw=none, processblue!40,
            error bars/.cd,
                y dir=both,
                y explicit relative,
        ] coordinates {          
            (RO,2.239) +- (0,0.01)
             (WO,2.343) +- (0,0.1)
             (EO,2.244) +- (0,0.02)
            (R/W,2.89) +- (0,0.03)
            (NO-ACC,2.234)  +- (0,0.1)
                        };

        \addplot+ [draw=none, blue!60,
            error bars/.cd,
                y dir=both,
                y explicit relative,
        ] coordinates {          
            (RO,2.91) +- (0,0.13)
             (WO,2.53) +- (0,0.11)
             (EO,2.58) +- (0,0.12)
            (R/W,2.96) +- (0,0.13)
            (NO-ACC,2.53)  +- (0,0.12)
                        };

                \addplot+ [draw=none, green!40,
            error bars/.cd,
                y dir=both,
                y explicit relative,
        ] coordinates {          
            (RO,2.932)  +- (0,0.13)
            (WO,2.609) +- (0,0.12)
            (EO,2.598) +- (0,0.11)
            (R/W,2.437) +- (0,0.13)
            (NO-ACC,2.026) +- (0,0.12)

                        };
        \end{scope}

\legend{mprotect,imc\_mprotect,hw imc\_mprotect}
    \end{axis}
\end{tikzpicture}}
\end{figure}

%% file: openssl.tex
   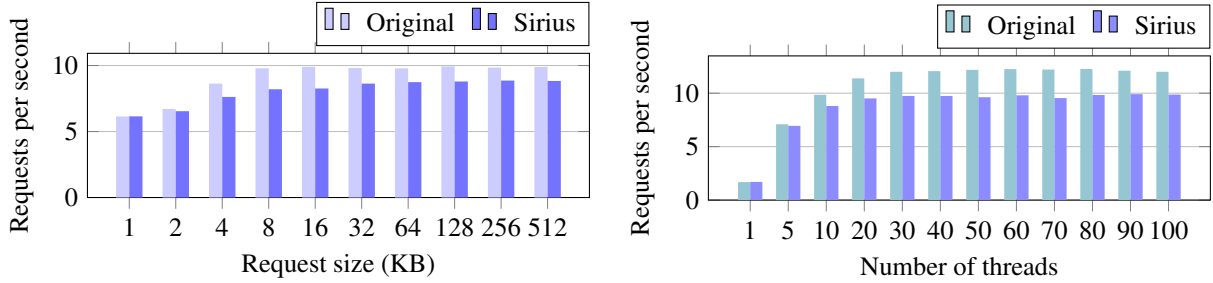
\begin{figure*}[t]
 \begin{subfigure}{.5\textwidth}
 \begin{tikzpicture}
    \begin{axis}[
        width=\textwidth,
        height=3.5cm,
        bar width=5pt,
        symbolic x coords={1,2,4,8,16,32,64,128,256, 512},
        xlabel={{Request size (KB)}},
        xtick=data,
        ymin=0,
        ybar=0pt,
        ymajorgrids=true,
        ylabel={{Requests per second}},
        legend cell align=left,
    legend style={
            at={(1,1.05)},
            anchor=south east,
            legend columns=3,
            column sep=1ex
        },
    ]
        \begin{scope}[
            draw=none,
        ]

\addplot  [draw=none,fill=blue!20  ] coordinates { (1,6.14) (2,6.71) (4,8.63) (8,9.79) (16,9.90) (32,9.81) (64,9.78) (128,9.93) (256,9.84) (512,9.89)  };
\addplot  [draw=none, fill=blue!55] coordinates {  (1,6.15) (2,6.54) (4,7.62) (8,8.20) (16,8.26) (32,8.63) (64,8.74) (128,8.79) (256,8.86) (512,8.83) };

        \end{scope}

\legend{Original,Sirius}
    \end{axis}
\end{tikzpicture}

\end{subfigure}\hfill
 \begin{subfigure}{.5\textwidth}

\begin{tikzpicture}
    \begin{axis}[
        width=\textwidth,
        height=3.5cm,
        bar width=4.6pt,
        symbolic x coords={1,5,10,20,30,40,50,60,70, 80, 90,100},
        xlabel={{Number of threads}} ,
        xtick=data,
        ymin=0,
        ybar=0pt,
        ymajorgrids=true,
        ylabel={ {Requests per second}},
        legend cell align=left,
    legend style={
            at={(1,1.05)},
            anchor=south east,
            legend columns=3,
            column sep=1ex
        },
    ]
        \begin{scope}[
            draw=none,
        ]

\addplot  [draw=none,fill=processblue!40 ] coordinates { (1,1.68) (5,7.09)    (10,9.85) (20,11.38) (30,12.00) (40,12.06) (50,12.18) (60,12.25) (70,12.21) (80,12.26) (90,12.10) (100,12.00)  };
\addplot  [draw=none, fill=blue!45] coordinates {  (1,1.7)(5,6.94)   (10,8.8) (20,9.51) (30,9.74) (40,9.74) (50,9.62) (60,9.8) (70,9.54) (80,9.83) (90,9.91) (100,9.87) };

        \end{scope}

\legend{Original,Sirius}
    \end{axis}
\end{tikzpicture}

\label{httpd}

\end{subfigure}
\caption{Overhead of Sirius-assisted httpd}
\label{fig:shttpd} 
 \end{figure*}

%% file: ml.tex
\begin{table}[htp]
\resizebox{.88\columnwidth}{!}{%

    \begin{tabular}{|l|c|c|c|}
    \hline
    Operation       &       {Baseline} & {Sirius} & {OPTEE}   \\ \hline
      training        & 75234 \textmu s & 78486 \textmu s  & 85354 \textmu s  \\ \hline
    pre-trained           &  68753 \textmu s & 73987 \textmu s  & 76453 \textmu s \\ \hline
    inference         &  33.23 \textmu s & 36.32 \textmu s & 38.45 \textmu s  \\ \hline

    \end{tabular}}
\end{table}

%% file: dds.tex
   \begin{figure}[htb]
\resizebox{\columnwidth}{!}{%

 \begin{tikzpicture}
    \begin{axis}[
        width=0.7\textwidth,
        height=4.5cm,
        bar width=7pt,
        symbolic x coords={handshake,DH,identity\_handle,identity\_cert,identity\_PK,session\_key,shared\_secret,aes\_ops,manage\_object,key\_material,challenges,canary},
        xlabel={},
        xtick=data,
        x tick label style={rotate=45,anchor=east},
        ymin=0,
        ybar=0pt,
        ymajorgrids=true,
        ylabel={Latency( ms)},
        ymode=log,
         legend cell align=left,
        legend style={
            at={(1,1.05)},
            anchor=south east,
            legend columns=2,
            column sep=1ex
        },
    ]
        \begin{scope}[
            draw=none,
        ]

        \addplot [draw=none, fill=pink!90] coordinates {
        (handshake,1274.56)
        (DH,558.12)
         (identity\_handle,1201.59)
        (identity\_cert,7213.195)
        (identity\_PK,2890.04)
        (session\_key,3790.57)
         (shared\_secret,51752)
        (aes\_ops,1885.7)
           (manage\_object,706.8)
        (key\_material,1369.14)
         (challenges,289.53)
        (canary,3.6)

        };
         \addplot [draw=none, fill=blue!45] coordinates {
        (handshake,1274.80) 
        (DH,576.44)
         (identity\_handle,1205.30)
        (identity\_cert,7216.4)
        (identity\_PK,2885.5) 
        (session\_key,3780.04) 
        (shared\_secret,51763)
        (aes\_ops,1887.75)
         (manage\_object,704.22)
        (key\_material,1373.35)
         (challenges,290.92)
        (canary,3.74)
        };

        \end{scope}

\legend{OP-TEE, Sirius }
    \end{axis}
\end{tikzpicture}}
\label{ddssec}
 \end{figure}

%% file: related.tex

\section{Related Work }\label{related}
Our goal with building Sirius has been to understand how to integrate TEE hardware to securely and pragmatically protect applications. Existing solutions are piecemeal (Table~\ref{cvetab}) but have not considered the need for system-wide compartmentalization and secure sharing for complex applications.

\tpara{ Partitioning frameworks} help developers to split applications to two trusted and untrusted components; such as Intel’s SGXSDK~\cite{intel2019linuxsgx}, Microsoft’s Open Enclave~\cite{microsoft2019openenclave}, Google’s Asylo~\cite{Google2018Asylo}, OP-TEE~\cite{optee}, and Keystone~\cite{dayeol2019keystone}. There are also language-specific partitioning frameworks such as Civet~\cite{tsai2020civet} for porting Java classes into an SGX enclave, TLR (Trusted Language Runtime)~\cite{santos2014using} for running portions of C\# applications inside TrustZone, and Glamdring~\cite{lind2017glamdring} a compiler for partitioning applications into SGX enclaves via code annotation.
Despite improving application security, none of these fully consider the complex attack surface originating from insecure interactions between kernel objects at both secure and normal worlds. 
Civet uses dynamic taint-tracking to control the flow of objects on enclave interfaces, but as we showed, data in TEE system flows within various kernel objects and not just interfaces between the two worlds (e.g., horizontal privilege escalation (HPE) attacks~\cite{suciu2020horizontal}). Sirius does not rely on a specific language~\cite{roy2009laminar} and can be used by these frameworks.
Also, targeting memory vulnerabilities requires in-address space memory protection, which is not considered in these works.

\tpara{ In-enclave LibOSs} focuses on porting unmodified applications entirely to enclaves~\cite{priebe2019sgx,baumann2015shielding,tsai2017graphene,arnautov2016scone}. For example, SGX-LKL~\cite{priebe2019sgx} ports a large part of the Linux kernel into an enclave and provides encrypted communication channels. However, this approach has a high overhead, limited compatibility for complex applications, and does not protect the host applications and OS from the enclave. Compromised or malicious third-party enclaves can collect and leak sensitive data ~\cite{schwarz2019practical,marschalek2018wolf} and transfer them through OS standard abstractions such as files or network sockets (Figure~\ref{fig:risks}~\ding{174}\ding{175}).
Existing OSs and TEE systems offer no comprehensive protection against such attacks. Additionally, such a large in-enclave codebase requires in-enclave compartmentalisation (i.e., supported by Sirius).
EnclaveDom~\cite{melara2019enclavedom} utilizes Intel MPK to provide in-enclave memory isolation, and MPTEE~\cite{zhao2020mptee} uses Intel MPX for providing protected shared memory.
Nested enclaves~\cite{park2020nested} modifies SGX hardware memory controller to enable multi-level security inside an enclave.
 Sirius is the first TEE compartmentalization framework that offers collaborative isolation at inter- and intra-address space levels without any hardware modification. It enables applications to architect various defense layers both in normal and enclave worlds. 

\tpara{ OS-assisted isolation:}
Providing isolation is historically one of the main jobs of OSs. The popular techniques include process-based isolation~\cite{bittau2008wedge,brumley2004privtrans},
namespaces~\cite{pike1992use,nsjail,gvisor,firejail}, capability-based sandboxing~\cite{shapiro2002eros,bomberger1992keykos,watson2012taste,loscocco2001integrating},   
mandatory access controls (MACs)~\cite{loscocco2001integrating,apparmor}, and system call filtering~\cite{seccomp}. Despite working well for protecting the host OS, they are not designed for pervasive compartmentalization and supporting mutually untrusting kernels running in different privilege levels, as is the case in the TEEs. Sirius is the first system to work alongside these in Linux.
SGXJail~\cite{weiser2019sgxjail} relies on process-based isolation and syscall filtering for sandboxing enclave malware. 
Sirius not only targets malware enclaves but also allows mutually distrustful compartments in each world. Its efficient IMC-based intra-process protection achieves significant performance improvement compared to inflexible process-based isolation and does not rely on specific hardware support ~\cite{belay2012dune,wangseimi,vahldiek2019erim,frassetto2018imix,tarkhani2020mu}. Previous DIFC systems~\cite{wang2015between, krohn2007information,zeldovich2008securing,zeldovich2008securing}, and in particular HiStar~\cite{zeldovich2006making}, inspired our work; However, these systems are not designed to achieve Sirius's goals for enabling compartmentalization in TEE systems on conventional OSs.



%% file: discussion.tex
\section{Discussion}\label{diss}

The consequences of the semantic gap between hardware privilege levels are relatively well studied in TEE systems that motivated us to build Sirius. However, other hardware security features such as Intel Trust Domain Extension (TDX)~\cite{tdx} and AMD SEV, will introduce a similar semantic gap in the hypervisor layer that are not explored yet, and we hope our work help to investigate the gap more and design a more secure systems stack. 
Our experience with building Sirius has shown a sweet spot for the adoption of DIFC to enable a unified framework for supporting mutually-distrustful compartments running inside different hardware privilege levels. However, this was not practical without~\rom{1} our kernel extensions to reduce the overhead of labeling within both kernels' abstractions, and in particular, within address space objects; and~\rom{1} our APIs to hide the underlying complexities that allow our ported applications to gain multiple layers of protection and very natural integration with Linux programming facilities. 

When porting Sirius applications, we learned DIFC works best when a clean definition of trust boundaries is possible. The right set of APIs that enables applications with simple compartmentalisation and seamless labeling, such as our extension to existing kernel syscalls, plays a significant role in adopting complex applications. In particular, in TEE systems, a typical target application already has a relatively clear sense of its secrets (for example, private keys or data models) to isolate inside enclaves, as well as its coarse-grained trusted and untrusted partitions; this is a perfect match for defining even more powerful trust boundaries.
That is why our ongoing work on enabling Sirius for SGX systems require only straightforward engineering and minor modifications in the IMC implementation (e.g., x86-64 has five layers of address translation instead of two in aarch32) and the SGX software stack. 
Still, the mechanism can be taken further---as modern CPUs support more features for fine-grained privilege separation such as memory tagging extension~\cite{mte} or hardware capabilities (e.g., CHERI~\cite{watson2015cheri}), Sirius can be adapted to support them.

%% file: conc.tex

\section{Conclusion}\label{conc}

We have presented Sirius, the first TEE-aware compartmentalisation framework to bridge the semantic gap between trusted and untrusted worlds. It enables secure sharing and strong multi-threaded privilege separation within both worlds by controlling dataflow across kernel objects. Sirius can guard complex applications against existing vulnerabilities and shows performance and security improvements in existing TEE-assisted applications.

%% file: appendix.tex
\appendix
\section{Appendix}\label{appendix:attacks}

\begin{table*}[t]
\centering
\renewcommand{\arraystretch}{.7}

    \begin{tabular}{|l|l|l|l|}
    \hline
    \textbf{Application name }                       & \textbf{Description}                                                & \textbf{TEE } & \textbf{Category }                               \\ \hline
    ltzvisor                                & TrustZone-assisted Hypervisor                              & TZ   & Reference monitor  \\ \hline
    LibSEAL                                 & SEcure Auditing Library for internet services              & SGX  & Auditing \\ \hline
    darknetz                                &  Darknet DNN framework                                     & TZ   & Data analytics                          \\ \hline
    sgx-spark                               & In-enclave Apache Spark                                    & SGX  & Data analytics                          \\ \hline
    shadow-box                              & Kernel Protector                                           & TZ   & Reference monitor \\ \hline
    SGX-Tor                                 & Tor anonymity network in the SGX                           & SGX  & Web app                                 \\ \hline
    TaLoS                                   & TLS Termination Inside Enclaves                            & SGX  & SSL/TLS                                 \\ \hline
    MQT-TZ                                  & TrustZone Enabled MQTT Broker                              & TZ   & Data analytics                          \\ \hline
    optee-sks                               & Library for Secure Key Services                            & TZ   & Key management                          \\ \hline
    Enclave EVM                             & Enclave Ethereum Virtual Machine                           & SGX  & Blockchain                              \\ \hline
    fabric-optee-chaincode                  & Hyperledger Fabric chaincode execution                     & TZ   & Blockchain                              \\ \hline
    fabric-private-chaincode                & In-enclave Chaincode Execution                             & SGX  & Blockchain                              \\ \hline
    self-healing\_FreeRTOS                  & A self-healing FreeRTOS                                    & TZ   & Reference monitor                       \\ \hline
    graphene-httpd                          & In-enclave httpd                                           & SGX  & Web apps                                \\ \hline
    graphene-nginx \& redis                 & In-enclave nginx \& redis                                  & SGX  & Web apps                                \\ \hline
    rustZone-backed-Bitcoin-Wallet          & An embedded Bitcoin wallet                                 & TZ   & Blockchain                              \\ \hline
    keyvault                                & Library for generating, storing and distribute secret keys & TZ   & Key management                          \\ \hline
    tzMon                                   & security framework for a mobile game application           & TZ   & Reference monitor \\ \hline
    SGX\_SQLite                             & SQLite database inside an enclave                          & SGX  & Databases                               \\ \hline
    sgx-lkl-MySQL                           & In-enclave MySQL                                           & SGX  & Databases                               \\ \hline
    SGX-OpenSSL                             & SGX SSL cryptographic library                              & SGX  & SSL/TLS                                 \\ \hline
    mbedtls-SGX                             & A SGX-friendly TLS stack                                   & SGX  & SSL/TLS                                 \\ \hline
    sgx-ra-tls                              & Integrate SGX remote attestation into the TLS              & SGX  & Attestation                             \\ \hline
    WolfSSL                                 & WolfSSL with SGX                                           & SGX  & SSL/TLS                                 \\ \hline
    slalom                                  & Private Execution of ML                                    & SGX  & Data analytics                          \\ \hline
    TresorSGX                               &  Securing storage encryption                               & SGX  & Databases                               \\ \hline
    SGX-LKL tensorflow \& pytorch & In-enclave tensorflow \& pytorch                 & SGX  & Data analytics                          \\ \hline
    stealthdb                               & Extension to PostgreSQL leveraging enclaves                & SGX  & Databases                               \\ \hline
    bolos-enclave                           & Ledger BOLOS Enclave                                       & SGX  & Blockchain                              \\ \hline
    Anonify                                 & A blockchain-agnostic execution environment                & SGX  & Blockchain                              \\ \hline
    SecureKeeper                            & In-enclave ZooKeeper                                       & SGX  & Web apps                                \\ \hline
    node-secureworker                       & In-enclave Java Scripts                                    & SGX  & Web apps                                \\ \hline
    SGX-Log                                 & Securing System Logs                                       & SGX  & Reference monitor  \\ \hline
    custos                                  & OS Auditing                                                & SGX  & Reference monitor  \\ \hline
    sgxjail                                 & Enclave sandboxing                                         & SGX  & Reference monitor  \\ \hline
    TEE-TLS-delegator                       & TLS sign delegator                                         & TZ   & SSL/TLS                                 \\ \hline
    SGX-pwd                                 & passwords distribution                                     & SGX  & Key management                          \\ \hline
    Panoply-Tor                             & In-enclave Tor                                             & SGX  & Web apps                                \\ \hline
    openenclave-tpm                         & Virtual tpm                                                & TZ   & Attestation                             \\ \hline
    \end{tabular}
    \caption{The full list of TEE-enabled applications that we studied.}
\end{table*}

\input{imc}

\input{mmstruct.tex}

\clearpage

%% file: imc.tex
\begin{lstlisting}[caption={Internal IMC's data structure },captionpos=b,label={imcstruct}]
struct imc_struct {
    //operation bitmaps: set to 1 if imc[i] 
    //is allowed to do this operation, 0 OW
    DECLARE_BITMAP(imc_Read, IMC_MAX); 
    DECLARE_BITMAP(imc_Write, IMC_MAX);
    DECLARE_BITMAP(imc_Execute, IMC_MAX); 
    DECLARE_BITMAP(imc_Allocate, IMC_MAX);
    int imc_id;    
    struct mutex imc_mutex;
    struct mem_segment *imc_range;
   };
\end{lstlisting}

%% file: mmstruct.tex
\begin{lstlisting}[caption={Modifications of mm\_struct },captionpos=b,label={mmstruct}]
struct mm_struct {
...
#ifdef CONFIG_SW_MCA
    struct imc_struct *imc_metadata[IMC_MAX];
    atomic_t num_imc;	/* number of imcs */
    /*imc Page tables per threads.*/
    pgd_t *imc_pgd_list[IMC_MAX];  
    int curr_using_imc;
    spinlock_t sl_imc[IMC_MAX];
    struct mutex imc_metadata_mut;
    DECLARE_BITMAP(imc_InUse, IMC_MAX);
#endif
... };
\end{lstlisting}